# Search of Secondary Pulsation Modes:

## Globular cluster *(NGC 6496)*

Gireesh C. Joshi and R. K. Tyagi

Department of physics

H. N. B. Govt. P. G. College, khatima

Udham Singh Nagar, India

gchandra.2012@rediffmail.com

*Abstract*— The Fourier-discrete-peridogram are used to identify pulsation modes in variables. We have found two pulsation modes in V1 and V2 among 13 new variables as described by Abbas et al.. The five variables V9 to V13 are not shown close to periodic values by analysis of the frequency distribution of multi-band data and also create difficulty to describe their varied nature. The multi-band periodic values of V1 and V6 are matched with known literature values. The scattering of the varied nature of secondary pulsation modes is eliminated by moving average methodology. The phase curve of secondary mode is found to be more smooth compared to a prominent mode of pulsation.

## INTRODUCTION

The clusters play an important role to understand the star formation history and galactic dynamics and evolution processes. These clusters of Milky Way are divided into open, globular and OB association groups, in which, the Globular clusters (GCs) become an ideal tool to study the properties and evolution of old stellar populations [1]. Bulge GCs are key templates of simple stellar populations for studying the stellar and chemical evolution in the high-metallicity domain [2]. Some stars of these populations change their brightness with time and called variables and such variability are found to be those stars having a stage of instability strips of stellar evolution. The theoretical models of stellar pulsation for observed stellar pulsation properties (periods, amplitudes, surface velocities, etc.) can provide additional constraints on the properties of stellar interiors and evolution [3]. These positions are used to divide variable stars into two groups such as short-periodic-variable (STV) and long-periodic-variable (LPV) stars. The STV is identified through whole night observations, whereas the detection of LPV is more challenging and time consuming process [3]. Abbas et. al., 2015 [2] have reported 13 new variable stars within GC NGC 6496, of which 6 are LPVs, 2 are suspected LPVs, and 5 are short-period eclipsing binaries. They have identified these variables using photometric data for 2-years. The NGC 6496 is a metal rich cluster, therefore, the variability of LPVs may also occur in either the motion of interstellar dust within cluster or in interaction between interstellar dust and stars. Eclipsing binaries are of interest in terms of stellar dynamics and stellar evolution in comparison of a single star. In addition, their period limit is still an open issue, and may need revision when more very short period systems are documented [4]. The detection of pulsation mode and analysis of pulsation frequencies provides a unique opportunity to understand their internal structure [5]. The variability of stars may be occurring due to many causes which further produced the multi-pulsation mode in variables. Pulsations lead the information about the physical properties of stars, including their masses, luminosities, temperatures and metallicities. Furthermore, pulsating stars cover a broad range of stellar parameters and evolutionary stages [6]. The analysis of observations of such type variables are important to infer about stellar parameters like their mass, radius and luminosity [7]. An identification of such pulsation is possible through Fourier-Discrete-periodogram (FDP) of variable stars. The improvement of phase coverage and quality of the various photometric light curves through Fourier decomposition methods would be effective to model and compare the light curves and evolution environment of variable stars [8]. The detail procedure of identification of secondary pulsation mode is described in the next section.

## METHODOLOGY OF SEARCH OF PULSATIONS MODE

The variation of brightness of stars with time is known as the light curve of variable. These light curves are used to examine the varied nature of stars. The light curve of regular variable is repeated after a constant time value, which is period of variable, whereas the irregular variables show sufficient amplitude variation but not repeated variation in a regular pattern. The Fourier transformation is utilized to estimate the peak amplitude and its corresponding frequency for regular variables, which leads to compute the periods of variability. The frequency-distribution (FD) of these variables is contained to be the multi secondary peaks (each having its own Gaussian distribution). The amplitude of these pulsations is higher than the background / noise FD but lower than the prominent peak value. The prominent peak value gives exact variable nature, but scattered variable nature is found through secondary peaks.

If the secondary peak does not belong to stellar variability, then the complete cycle of variability does not occur. As a result, the possibility of false estimation of secondary peak is denied. After deep investigation, we have found secondary pulsation modes in 5 variables among 13 variables as identified by Abbas et al., 2015 [3]. The above described frequency distribution is found using "PERIOD04" software. The period values of variables through different bands are listed in Table 1. The period values of variables through various bands are not found similar to each other except V1 and V6. These results are occurring due to the very high field decontamination and reddening in GC. These results supported an idea that the variable content is invaluable for better determination of physical parameters [9].

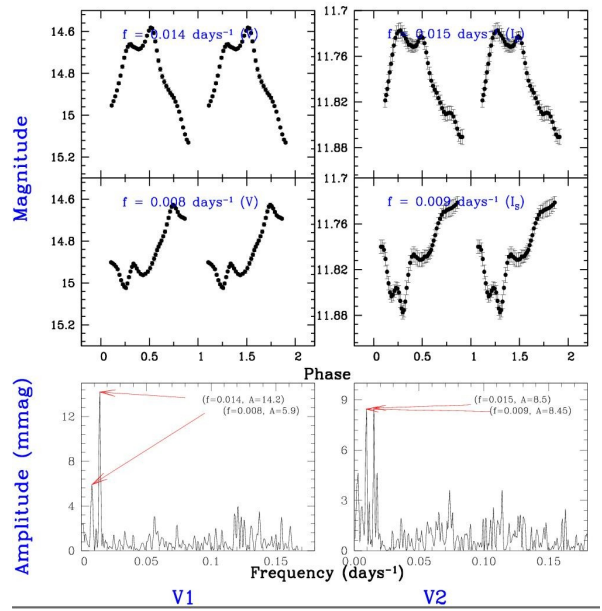

*Figure 01. (The frequencies of pulsation mode are shown in lower panels whereas upper and middle panels are represented the possible phase diagrams through moving average method.)*

*Table 1: The period values of variables in different pass-bands.*

| Variable Name | Period by Abbas et al. (2015)[1] | Periods in different Pass-bands (In days) | | |
|---|---|---|---|---|
| | | $V$ | $I_L$ | $I_S$ |
| V1 | 69 | 70.133 | 70.721 | 71.047 |
| V2 | 55 | 0.014 | -- | 65.438 |
| V3 | 37 | 0.997 | 0.997 | 355.238 |
| V4 | 73 | 1.008 | -- | 75.353 |
| V5 | 85 | 95.217 | -- | 95.641 |
| V6 | 48 | 47.501 | -- | 46.918 |
| V7 | -- | 32.854 | 0.061 | -- |
| V8 | -- | 80.189 | 0.081 | -- |
| V9 | 0.766 | 0.074 | 0.383 | 4.326 |
| V10 | 0.289 | 0.144 | 0.144 | 10.536 |
| V11 | 0.337 | 415.110 | 0.336 | 0.168 |
| V12 | 0.884 | 0.327 | 0.442 | 4.713 |
| V13 | 0.438 | 0.059 | 0.219 | 248.667 |

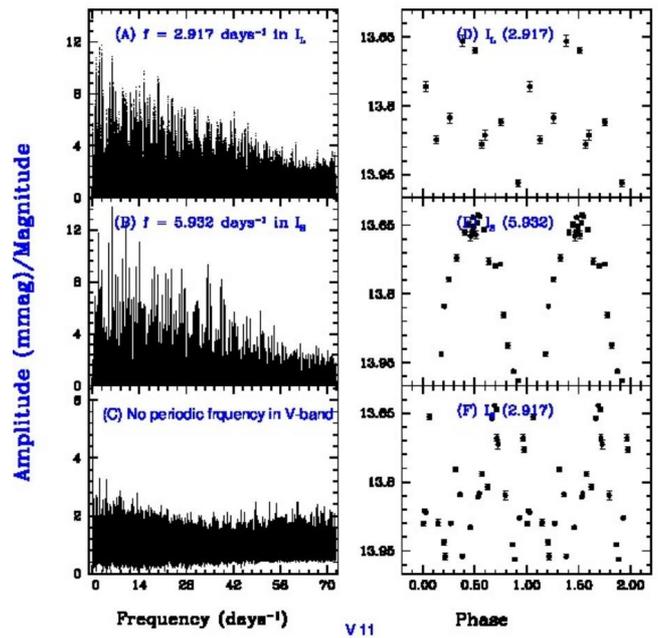

*Figure 02. (The amplitude variation with frequency-distribution of pulsation star are shown in lower panels, whereas right panels are represented the corresponding phase diagrams of V11 through moving average method.)*

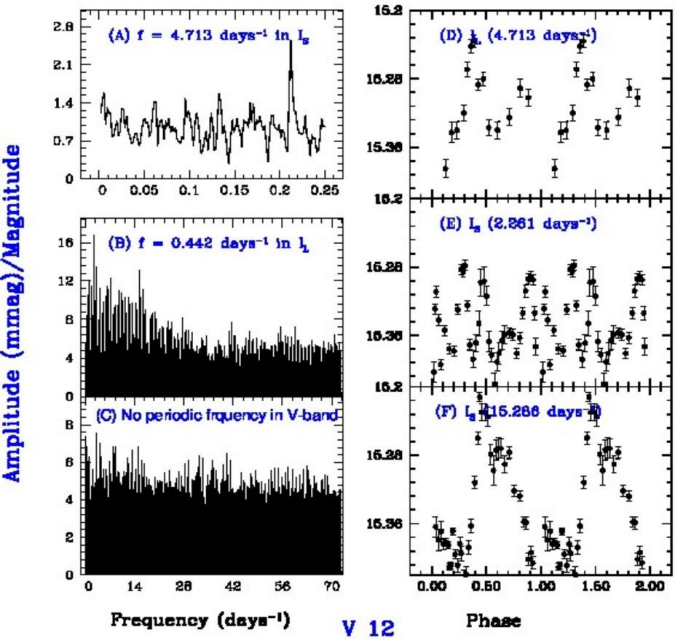

Figure 03. *(The frequencies of pulsation mode are shown in left panels, whereas right panels are represented the possible phase diagrams in various pulsation mode of V12.)*

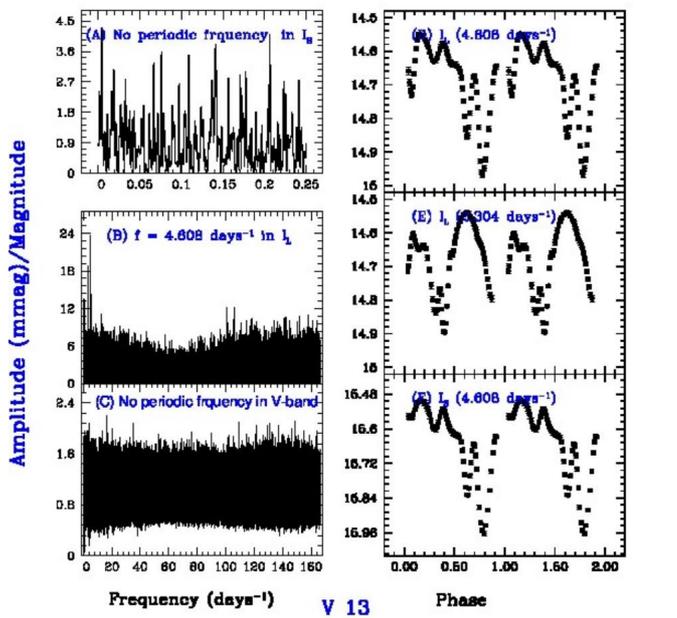

Figure 04. *(The frequencies of pulsation mode are shown in left panels, whereas right panels are represented the possible phase diagrams in various pulsation mode of V13.)*

CONSTRUCTION OF light-folded CURVE

The light folded curves of variables, i.e., phase diagram are constructed through the iteration procedure of moving average. In this procedure, the phase points of each variable have been computed for each data point by assuming first epoch time as a reference point. This reference point is altered according to the first time point of datasets in various bands. These data points are arranged in increasing order according to their phase values. After this, the shortest five points are selected and their average is determined. This average value is provided the first data point of the new dataset of the phase. The shortest data point from the old phase dataset is eliminated and new five shorter points from remaining data points are selected. The next value of phase data point is computed through these data points. This procedure is repeated until we are not computed the average of last remaining five data points. The new phase data points are one less than that of old data-points of phase. The variability scattering is reduced through this procedure, but remaining scattering is too high for checking the variable nature, therefore, he whole methodology has been repeated four times. The variability curve is found to be smooth, but it has four less points in comparison to original data-points of phase.

DISCUSSION

The Gaussian distribution of amplitude and frequency are not found for any variable within present studied GC. This may be an effect of fewer data points. The lack of data-points reduces the efficiency of searching secondary modes of pulsations. We have found two-two pulsation modes in V1 and V2 variables as shown in the lower panels of Figure 01. The phase diagrams of these possible modes are shown in upper and middle panels of Figure 01. The gap in phase arises due to the reduction of data-points through moving average method. Although these variables are LPV and their corresponding amplitude peaks appeared in the frequency periodogram. In these diagrams, the peak values are suddenly occurring, but too high for background noise effect. The background noise effect produces the amplitude peaks in the theoretical FD and the exact corresponding variability nature of these peaks confirms them as a secondary pulsation modes. Similar way, we have found two period values for V11 as 0.336 and 0.168 days using the data set of Abbas et al .(2015) in $I_l$ and $I_s$ magnitudes of stars. Although, one periodic value of this variable (0.337 days) is also approximately found in our present result. On the other hand, no period value is accessible through dataset of V-band of V11. The varied nature is found to be smoother for 0.168 days compared to that of 0.336 days. It is an interesting thing that the varied nature is found to be smoother for secondary pulsation frequency rather than prominent variable frequency. Such results are motivated to use for searching the secondary pulsation mode in variable stars. The frequency analysis of V12 is showing different peak values in different bands, which creates conflict to describe the varied nature, however, the frequency 15.286/days clearly leads a smooth variable curve compared to others. On analysising the amplitude-frequency diagram of V13, the peak value of amplitude is found at 4.608/days of $I_L$. The shape of the phase-light curve of this variable indicates that it may be belong to the class of eclipsing binary variables. The effective frequency of such type variables assumes to be half that of computed frequency (as shown in Figure 4). There are no frequency peaks of V13 in V and $I_S$ bands.

## CONCLUSION

The peak value of amplitude in Lomb-Scargle normalized periodogram (frequency distribution) of variables is not itself sufficient for having a pulsed mode of variable. The procedure of secondary pulsation search in variable stars provides a unique opportunity to identify secondary pulsation modes in variable stars. The secondary pulsation modes are low from principle peak value, but too low for merging with that background noise frequency, these noise frequencies are arising due to either amplitude noise or photometric scattering. The results of different periods of different photometric bands of varying is indication of various physical phenomena within variables and the radiating energy of these processes becomes in prominent in different wavelengths. This study is suggested to search variability of variables in different photometric bands and construct the possible phenomena for describing these pulsation modes. Furthermore, the shift in pulsation mode of variable in bands provide to clue for searching such type effect.

## RESULTS

We have been finding different period values for the variables compared to that of Abbas et al. (2015)[3]. The period values of V1 and V6 are found to be close to each other either obtained by photometric data of various bands or as obtained by Abbas et al. (2015)[3]. These authors are reporting that all stars from V9 to V13 are short-period variables (EW contact and EB ellipsoidal eclipsing binaries). However, our present results of periods of these variables are not found similar in multi-band analysis. We have compared our results with known previous results as prescribed by Abbas et al. (2015)[3]. The smoothness of the phase-light curve of variables is increased through the moving average method and this method deals to constrain the importance of searching of secondary pulsation modes. Furthermore, the variability curve of some variables is smoother for these modes rather than the peak amplitude of frequency. In this background, their type is doubtful and shows need of all night observations of these variables. The amplitude of secondary modes is found to be less than that of prominent peak amplitude.

## ACKNOWLEDGMENT

This research has made use of the VizieR catalogue access tool, CDS, Strasbourg, France. The original description of the VizieR service was published in A&AS 143, 23. GCJ is thankful to APcyber Zone (Nanakmatta) for providing computer facilities.